\documentstyle[12pt]{article}
\begin{document}
\hfill hep-th/9705101
\begin{center}\Large
New Constraints on the Yukawa-Type Hypothetical Interaction From The Recent
Casimir Force Measurement \end{center}

M. Bordag\footnote{Institute for Theoretical Physics, Leipzig
University, Augustusplatz 10/11, 04109 Leipzig, Germany,
http://www.physik.uni-leipzig.de/~bordag},
G.T. Gillies\footnote{Department of Physics, University of Virginia,
Charlottesville, Virginia 22901, email: gtg@faraday.clas.virginia.edu}
and
V.M. Mostepanenko\footnote{Department of Physics, Federal University
of Paraiba, C.P. 5008, CEP 58059-970, Joao Pessoa, Pb-Brazil,
e-mail: mostep@dfjp.ufpb.br (on leave from A.Friedmann Laboratory
for Theoretical Physics, St.Petersburg, Russia)}

\begin{abstract}
  We calculate the constraints on the constants of hypothetical long-range
  interactions which follow from the recent measurement of the Casimir
  force. A comparison with previous constraints is given. The new
  constraints are up to a factor of 3000
stronger in some parameter regions .
\end{abstract}

\section{Introduction}

The existence of Yukawa-type long-range interactions between the atoms
of macro-bodies is predicted by the modern unified gauge-field
theories of fundamental interactions, from the principles of
supersymmetry and supergravity.  These forces may be understood as a
result of the exchange of light elementary particles (such as the axion,
scalar axion, dilaton, spin-one antigraviton etc. \cite{1}). They may
be called {\it long-range } because their range of action is much
larger than the typical nuclear sizes. Additional Yukawa-type
interactions may arise also from corrections to the Newtonian
gravitational law at small distances \cite{2,3}. At present, this
subject is actively investigated. There are a large number of papers
from both theoreticians and experimentalists. An extended annotative
collection of references can be found in \cite{4}.

The most reliable way to get constraints on the parameters of the
Yukawa-type long-range interactions is via laboratory experiments
which are sensitive to the presence of additional forces.  Examples
are the E\"{o}tv\"{o}s-, Galileo- and Cavendish-type experiments,
Casimir and van der Waals force measurements, and measurements of
transition probabilities in exotic atoms.

A collection of modern constraints following from the above-mentioned
experiments can be found in \cite{5}.

Casimir forces which arise between closely located surfaces of solid
bodies due to the presence of electromagnetic vacuum fluctuations
provide us with one of the most interesting possibilities to restrict
the permissable strength of the hypothetical interaction. A detailed
analysis of the Casimir effect, including those aspect under
discussion here can be found in the monograph \cite{6}.

It was first suggested in \cite{7} that constraints on the
Yukawa-type interactions can be derived from the Casimir effect.  The
strongest constraints, which follow from the Casimir force measurement
between a quartz plate and a spherical lens, were obtained in
\cite{5,8} (see also \cite{6}). They provide  the strongest
constraints for a range $\lambda$ of the Yukawa forces within
$10^{-8}{\rm m}\le\lambda\le 10^{-4}{\rm m}$. For $\lambda\ge
10^{-4}{\rm m}$ the best constraints followed from the Cavendish- and
E\"{o}tv\"{o}s-type experiments \cite{5,8} and for $\lambda\le
10^{-8}{\rm m}$ from van der Waals force measurements and from
measurements of transition probabilities in exotic atoms \cite{9}.
Therefore new, more exact Casimir force measurements are of prime
importance in order to obtain stronger constraints on the constants of
Yukawa-type hypothetical interactions in the intermediate range of
$\lambda$.

Recently, a new experiment was performed \cite{10}, in which the
Casimir force was measured between two metallised surfaces, a plane
disc and a spherical lens with a distance $a$ between them. The
theoretical value, obtained in \cite{11}, was confirmed within an
accuracy of 5\% in the distance region $0.6\mu{\rm m}\le a\le 6\mu{\rm
  m}$. Note that this was the first direct measurement of the Casimir
force for metallic surfaces since Sparnaay \cite{spaarney}, and that
the accuracy has been increased in 20 times.  The metallic surface of
the bodies (a $0.5\mu{\rm m}$ thick gold layer) increased considerably
the contribution of the hypothetical Yukawa forces due to the high
density of gold (the Yukawa forces are proportional to the density
squared).  Also, the relatively large spacing between the bodies, up
to $6 \mu{\rm m}$, where the Casimir force is very small, turned out
to be essential for the strengthening of the constraints.

In the present paper new constraints on parameters of the Yukawa-type
hypothetical interactions, which follow from the experiment \cite{10}
on Casimir force measurement, are obtained. They turn out to be
stronger than those obtained previously from the measurement of the
Casimir force between dielectrics and from one of the versions of a
Cavendish-type experiment in the range $6.3~10^{-8}{\rm
  m}\le\lambda\le 3.1~10^{-3}{\rm m}$. Especially within
$2~10^{-6}{\rm m}\le\lambda\le 3~10^{-5}{\rm m}$, they are stronger
than the previously known ones by about $10^{3}$ times and near
$\lambda\approx 6~10^{-6}$, they are $3~10^3$ times stronger.  The
corresponding constraints which result for the masses of the spin-one
antigraviton and for the dilaton are also discussed.

\section{Calculation of the Hypothetical Forces for the Configuration
Used in the Experiment}

In the recent experiment \cite{10} the Casimir force was measured
between a plane disc of diameter $d=2.54{\rm ~cm}$ and thickness
$D=0.5{\rm ~cm}$ and a lens of height $H=0.18{\rm ~cm}$, diameter
$L=4{\rm ~cm}$ and curvature radius $R=11.3{\rm ~cm}$. The surfaces of
the plate and the lens were covered with copper and gold layers of
$0.5\mu{\rm m}$ thickness each (see Fig. 1). For this configuration
the Casimir force is attractive and given by \cite{11}
\begin{equation} F=-{2\pi R\over 3} {\pi^2\over 240} {\hbar{\rm
      c}\over a^3}~.\label{1}\end{equation}
This expression
was later obtained by another method in Ref. \cite{12} (see also Refs.
\cite{6,13}). In fact it is an approximation but, as shown in
\cite{6,13}, for a large radius-of-curvature   lens, the relative error
is less than 0.01\%.

The potential of the hypothetical interaction between the atoms of the
disc and the atoms in the lens has the form \begin{equation}
V(r)=-\alpha\hbar{\rm c}N_{1}N_{2}{\exp(-r/\lambda)\over r}~,
\label{2}\end{equation} where $r$ is the interatomic distance,
$\alpha$ is the (dimensionless) interaction constant,
$\lambda=\hbar/{\rm mc}$ is the range of the interaction and
simultaneously the Compton wavelength of the hypothetical particles
whose exchange leads to the hypothetical forces under discussion. The
coefficients $N_{1},N_{2}$ are the numbers of nucleons in the atoms so
that the constant $\alpha$ is independent of the type of atoms, i.e.,
on the nuclear charges.

Due to its smallness the hypothetical interaction between solid
bodies can be obtained by summing the potential (\ref{2}) over their
volumes \begin{eqnarray} V(a)&=&-\alpha{\hbar{\rm c}\over m_{\rm
p}^{2}} \left [ \rho_{1}\int_{V_{1}}{\rm d^{3}}{\bf r}_{1}+
  \rho_{2}\int_{V_{2}}{\rm d^{3}}{\bf r}_{1}+  \rho_{3}\int_{V_{3}}{\rm
  d^{3}}{\bf r}_{1}\right ]\nonumber \\ &&
\times\left[  \tilde{\rho}_{1}\int_{\tilde{V}_{1}}{\rm d^{3}}{\bf r}_{2}+
  \rho_{2}\int_{\tilde{V}_{2}}{\rm d^{3}}{\bf r}_{2}+
  \rho_{3}\int_{\tilde{V}_{3}}{\rm d^{3}}{\bf r}_{2}\right ]
{\exp(-r/\lambda)\over r}\;, \label{3}\end{eqnarray} where $ r=|{{\bf
r}_{1}-{\bf r}_{2}}| $. The vectors ${\bf r}_{1}$ and ${\bf r}_{2}$
belong to the volumes indicated in the lower integration limits in the
operator brackets. This formula reflects the composition of the disc
and the lens, both having metal layers, used in the experiment. In
(\ref{3}), by means of $\rho=nNm_{\rm p}$, where $m_{\rm P}$ is the proton
mass, we introduced the densities of the materials used in the
experiment: $\rho_{1}$ and $\tilde{\rho_{1}}$ for the quartz in the
disc and the lens, respectively and $\rho_{2,3}$ for the copper
and gold layers ($n$ being the {\it atomic} densities).

Using this formula, the hypothetical force between the disc and the
lens can be expressed as \begin{equation} F_{H}=-{\partial V(a)\over
\partial a}~.  \label{4}\end{equation}

The integrals in these expressions cannot be taken
explicitely. In general,  numerical integration is required
(see below). However, if the part of the lens, which contributes most to
the result is small compared with the disc, the integrals
in (\ref{3}) can easily be taken explicitely. This approximation is in
any case justified for $\lambda\le 10\mu{\rm m}$. In fact, its validity
can be extended up to $\lambda{<\atop\sim}100\mu{\rm
  m}$ as the numerical calculations show.

In order to perform the calculation for an infinite disc we first
consider the potential energy of one lens atom having $N_L$ nucleons
in its nucleus at position $(0,0,l)$ above the disc so that the
distance to an atom within the disc at position $(x,y,z)$ is
$r=\sqrt{x^{2}+y^{2}+(z-l)^{2}}$  (see Fig. 1)
\begin{eqnarray}
V_{A}(l)&=&-\alpha N_L{\hbar{\rm c}\over m_{\rm
p}}\int_{-\infty}^{\infty}{\rm d}x\int_{-\infty}^{\infty}{\rm d}y
\left[ \rho_{1}\int_{-D}^{-2\Delta}{\rm d}z+
\rho_{2}\int_{-2\Delta}^{-\Delta}{\rm d}z\right.\nonumber\\
&&\left.~~~~~~~~~~~~~~~~~~~~~~~+
\rho_{3}\int_{-\Delta}^{0}{\rm d}z\right]{\exp(-r/\lambda)\over r}\,.
\label{6}\end{eqnarray} Performing the integration in (\ref{6}) we get
\begin{equation} V_{A}(l)=KN_{L}\exp(-l/\lambda)\;,
\label{7}\end{equation} where the coefficient $K$ is given by
\begin{eqnarray} K&=&-2\pi\alpha\lambda^{2}{\hbar{\rm c}\over m_{\rm
p}}
\left[\left(\rho_{1}-\rho_{2}\right)\exp(-2\Delta/\lambda)\right.\nonumber\\
&&\left.+(\rho_{2}-\rho_{3})\exp(-\Delta/\lambda)-\rho_{1}\exp(-D/\lambda)
+\rho_{3}\right]\;.  \label{8}\end{eqnarray}
In order to get the
complete interaction potential it is necessary to integrate (\ref{7})
over the volume of the lens, multiplying by the corresponding atomic
densities at height $l$. We obtain
\begin{eqnarray} V(a)&=&{K\over
m_{\rm
p}}\left\{\int_{a+2\Delta}^{a+H}\left[\gamma_{1}(l)+\gamma_{2}^{(1)}(l)+
\gamma_{3}^{(1)}(l)\right]\exp(-l/\lambda){\rm d}l\right.\nonumber\\
&&~~~~~~~+\int_{a+\Delta}^{a+2\Delta}\left[\gamma_{2}^{(2)}(l)+
\gamma_{3}^{(1)}(l)\right]\exp(-l/\lambda){\rm d}l\nonumber\\
&&~~~~~~~+\left. \int_{a}^{a+\Delta}\gamma_{3}^{(2)}(l)\exp(-l/\lambda){\rm
d}l\right\}\;, \label{9}\end{eqnarray}
where the $\gamma$'s are in
fact proportional to the areas of sections of the lens at height $l$
over the disc. They are given by:
\begin{equation} \gamma_{1}(l)=\pi
\tilde{\rho_{1}}\left[2R(l-a-2\Delta)-(l-a)^{2}+
4\Delta^{2}\right]\label{10}\end{equation} (for the quartz atoms
within $a+2\Delta\le l\le a+H$), \begin{equation}
\gamma_{2}^{(1)}(l)=\pi
{\rho_{2}}\Delta(2R-3\Delta)\label{11}\end{equation} (for the copper
atoms within $a+2\Delta\le l\le a+H$), \begin{equation}
\gamma_{3}^{(1)}(l)=\pi
{\rho_{3}}\Delta(2R-\Delta)\label{12}\end{equation} (for the gold
atoms within $a+\Delta\le l\le a+H$), \begin{equation}
\gamma_{2}^{(2)}(l)=\pi{\rho_{2}}\left[2R(l-a-\Delta)-(l-a)^{2}+
\Delta^{2}\right]\label{13}\end{equation} (for the copper atoms within
$a+\Delta\le l\le a+2\Delta$), \begin{equation}
\gamma_{3}^{(2)}(l)=\pi{\rho_{3}}
\left[2R(l-a)-(l-a)^{2}\right]\label{14}\end{equation} (for the gold
atoms within $a\le l\le a+\Delta$).

Performing the integration in (\ref{9}) we obtain the final expression
for the force $F_H^{inf}$ between an infinite disc and the lens:
\begin{eqnarray}
 F_H^{inf}&=&-2\pi^2\alpha{\lambda^2\hbar{\rm c}\over m_{\rm p}^2} {\rm
e}^{-a/\lambda}\left[\rho_3-(\rho_2-\rho_1){\rm e}^{-2\Delta/\lambda}-
(\rho_3-\rho_2){\rm e}^{-\Delta/\lambda}-\rho_1{\rm
e}^{-D/\lambda}\right]\nonumber\\
&&\times\Big\{2\lambda\rho_3(R-\lambda)-2\lambda(\rho_3-\rho_2)
(R-\Delta-\lambda){\rm e}^{-\Delta/\lambda} \nonumber\\ &&-2\lambda
(\rho_2-\tilde{\rho_1}) (R-2\Delta-\lambda){\rm e}^{-2\Delta/\lambda}
\nonumber\\ && +\left[2R\lambda\tilde{\rho_1}\left(-1+{\lambda/ R}-{H
/ \lambda}+{H/ R}+ {H^2/ 2R\lambda}\right) \right.\nonumber\\
&&\left.+\Delta^2(\rho_3+3\rho_2-4\tilde{\rho_1})-
2R\Delta(\rho_2+\rho_3-2\tilde{\rho_1})\right]{\rm
e}^{-H/\lambda}\Big\}\;.  \label{15}\end{eqnarray}

\section{Constraints for the Parameters of the Yukawa-Type Interaction}

The expression (\ref{1}) for the Casimir force acting between a flat
plate and a spherical lens was confirmed experimentally in \cite{10}
for a range of distance $0.6\mu{\rm m} \le a \le 6\mu{\rm m}$ with a
relative error of $\delta =5\%$. This means that the value of the
hypothetical forces cannot surpass five percents of the Casimir
force. Consequently the inequality
\begin{equation}
  \label{16}
  |F_H|\le 0.05|F|\;,
\end{equation}
where $F_H$ is given by Eqs. (\ref{3}), (\ref{4}) or, in the case of
sufficiently small $\lambda$ by Eq. (\ref{15}), holds.

From the inequality (\ref{16}) the constraints on the parameters
$\alpha$ and $\lambda$ of the hypothetical force $F_H$ follow. We
consider this in what follows.  Note that for the range $\lambda\le
10^{-7}{\rm m}$ the force is well approximated by the simple formula
following from (\ref{15})
\begin{equation}  \label{17}
  F_H^{inf}=-4\pi^2\alpha {\lambda^3\hbar {\rm c}R\over m_{\rm
p}^2}\rho_3^2{\rm e}^{-a/\lambda}\,.  \end{equation} Using (\ref{1}),
(\ref{17}), the inequality (\ref{16})
 may be written in the form
 \begin{equation}
   \label{18}
   |\alpha|\le {5\pi\over 1.44 ~10^5}{m_{\rm p}^2\over
\rho_3^2\lambda^3}{1\over a^3}{\rm e}^{a/\lambda}\,.
 \end{equation}
It is seen from (\ref{18}) that the strength of the constraint in this
range does not depend on $R$, the radius of the lens.

For the range $10^{-8}{\rm m}\le\lambda\le 1.6~10^{-4}{\rm m}$ the
constraints are obtained using formula (\ref{15}). For the range
$1.6~10^{-4}{\rm m}\le\lambda\le 10^{2}{\rm m}$ it is impossible to
consider the radius of the disc as large and a numerical integration
in formula (\ref{3}) was performed. For this purpose  algorithm
698 from netlib (http://www.netlib.org) was used.  It is an
adaptive multidimensional integration, the Fortran program is called
'dcuhre'.  A large number of function calls (about $5~10^6$) for
$\lambda\ge10^{-4}{\rm m}$ was necessary in order to obtain reliable
numerical values. For smaller $\lambda$ the program does not work, but
there the radius of the disc becomes unimportant and formula
(\ref{15}) can be used. Note that, e.g., for $\lambda=10^{-3}$m the
hypothetical force acting between the lens and the finite disc is
72.3\% of the force for an infinite disc. For $\lambda=10^{-4}$m it is
already 99.96\%.

The values of the geometrical parameters used
in \cite{10} are given in Sec. 2. The densities of the materials of
the disc and of the lens are $\rho_1=2.4 {\rm g}/{\rm cm}^3$ for the
quartz in the lens and $\tilde{\rho_1}=2.23 {\rm g}/{\rm cm}^3$ for
the quartz in the disc, $\rho_2=8.96{\rm g}/{\rm cm}^3$
and $\rho_3=19.32{\rm g}/{\rm cm}^3$ for copper and gold,
correspondingly.

The results of all calculations are collected into inequality
(\ref{16}) in order to obtain the new constraints on the constants of
hypothetical long-range interactions. The obtained constraints are
shown by curve 1 in Fig. 2, where the region above the curve is ruled
out by the cited experimental data. Thereby for every value of
$\lambda$ the value of $a$ was used which gives the best constraint.
This is $a=6\mu{\rm m}$ for large $\lambda$ and $a=0.6\mu{\rm m}$ for
small $\lambda$ and correspondingly in between. For comparison, in the
same Fig. 2 the curve 2 shows the previously known \cite{5,6,8}
constraints for the Yukawa-type hypothetical interactions which follow
from the experiment \cite{11} on Casimir force measuring. In that
experiment the lens and the disc (it was large as compared with the
lens) were made out from quartz only (without metallic layers on the
surfaces). The dependence (\ref{1}) with an appropriate correction
factor due to the dielectricity was confirmed with a relative error
of $\delta=10\%$ for distances $0.1\mu{\rm m}\le a \le 1\mu{\rm
m}$. As a result, for extremely small $\lambda\approx 10^{-8}{\rm m}$
the experiment \cite{11} gave stronger constraints than the experiment
\cite{10}. This constraint results from the smallest values $a=0.1
\mu{\rm m}$. This becomes clear when taking into account that the
function $a^{-3}\exp(a/\lambda)$ appearing in (\ref{18}) takes its
minimal value for $a=3\lambda$. Therefore the experiment \cite{11}
with the smallest value of $a$ gives the best results in this case.

For $\lambda\ge 6.3~10^{-8}{\rm m}$, as  can be seen from Fig. 2,
the experiment under consideration gives much stronger constraints
than that of \cite{11}. In the range $2~10^{-6}{\rm
m}\le\lambda\le 3~10^{-5}{\rm m}$ the new constraints surpass the old
ones by  more than a factor of 1000 and near $\lambda=6~10^{-6}$, by
3000.  For $\lambda\approx 10^{-4}{\rm m}$, where the old Casimir
experiment has given the same constraints as one of the Cavendish-type
experiments, the increase is  400 times.

 The constraints which follow from the Cavendish- and
E\"{o}tv\"{o}s-type experiments are also shown in Fig. 2
\cite{5,6,8}. Curve 3 follows from the experiment \cite{14}, curve 4
from \cite{15} and curve 5 from \cite{16}, curve 6 from the
E\"{o}tv\"{o}s-type experiment \cite{17}. As shown in Fig. 2, in the
range $10^{-4}{\rm m}\le\lambda\le 3.1~10^{-3}{\rm m}$ the new
constraints from the Casimir effect surpass in some points the results
from the Cavendish-type experiment in Ref. \cite{13}, in some points
they are of the same strength (compare the curves 1 and 3 in Fig. 2).

The new constraints resulting from the recent Casimir force
measurement (in distinction from the old ones) give the possibility to
get constraints for the masses of the spin-one antigraviton and the
dilaton. The exchange of spin-one antigravitons gives rise to
repulsive forces described by (\ref{1}) with $\alpha_a=-8\pi{\rm
G}{\rm m}_{\rm qe}^2/(\hbar{\rm c})$, where ${\rm m}_{\rm qe}\approx
365\rm MeV$ is the sum of the masses of the current quarks belonging
to a nucleus and of atomic electrons, and G is the gravitational constant.
This gives $\alpha\approx 10^{-40}$ and from  curve 1 of Fig. 2 it
follows that $\lambda_a\le 5~10^{-3}\rm m$, or for the spin-one
antigraviton mass, $m_a\ge 3.6~10^{-5}\rm eV$. This constraint is
only slightly weaker than the previous one  ($\lambda_a\le
3~10^{-3}\rm m$, $m_a\ge 6~10^{-5}\rm eV$, \cite{5}) obtained
from the Cavendish-type experiment of Ref. \cite{15} (shown as curve 4
in Fig. 2).

A similar constraint follows for the dilaton whose interaction
constant is given by  theory to be $\alpha_d=\frac{1}{3}{Gm_{\rm
    p}^2/ ( \hbar \rm c)}\approx 2~10^{-39}$. As  seen from Fig.  2,
for such a constant $\lambda_d\le 5.5~10^{-4}\rm m$ or, in terms of
the dilaton mass, $m_d\ge 3.3~10^{-4}\rm eV$ follows. Quite the same
constraint follows from the Cavendish-type experiment of Ref.
\cite{14} (shown as curve 3 in Fig. 2). Therefore in the region
$10^{-4}{\rm m}\le\lambda_d\le 3.1~10^{-3}\rm m$ the constraints
following from the Casimir effect are approximately of the same
strength as those which follow from Cavendish-type experiments.

\section{Conclusion and Discussion}

As discussed in the preceeding sections, the new measurement of the
Casimir force presented in \cite{10} makes it possibile to
strengthen the constraints of the constants of hypothetical
long-range interactions by a factor of 1000 and more for a wide
interaction range.  Moreover, it it is also possible
to widen (by one and a half orders of magnitude) the  boundary
of the $\lambda$-interval for which the Casimir effect gives
good constraints for  Yukawa-type hypothetical interactions.  In
this way the Casimir effect becomes a strong competitor to the
Cavendish-type experiments in obtaining constraints for such interactions.

 To obtain even stronger constraints from the Casimir effect
it would be desirable to enlarge the distance interval over which the force
is measured, perhaps up to  $0.1\mu{\rm m}\le a \le 10
\mu{\rm m}$. Near the left boundary of this interval,
one must take into account
the contributions to the Casimir force
resulting from the uneveness and imperfections of the disc and the
lens \cite{13}. Near the right boundary, the contributions due to  the
temperature corrections \cite{6} should be considered too. Also, the
use of materials with high density for the whole disc and lens would be
helpful. Finally, a further reduction of the error in the force
measurement would be important.

\section*{Acknowledgments}

The authors are greatly indebted to S.K. Lamoreaux for sending the
manuscript of his paper \cite{10} before publication and for several
helpful discussions, in particular for communicating the densities
$\rho_1$ and $\tilde{\rho_1}$. They also thank A.J. Sanders for his
interest in this work and for facilitating communication.

One of us (V.M.M.) is indebted to the Institute for Theoretical Physics
of the University of Leipzig, where this work was partly performed, for
kind hospitality and to the DFG under grant 436 RUS 17/88/94 for support.

\newpage
\section*{List of Captions}

Fig. 1 \parbox[t]{13cm}{The configuration of a flat disc with diameter
  $d$ and thickness $D$, and a spherical lens with curvature radius
  $R$, height $h$ and diameter $L$ used in the experimental
  measurement of the Casimir force \cite{10}. The volumes $V_1$ and
  $\tilde{V_1}$ are occupied by quartz, $V_2,\tilde{V_2}$ and
  $V_3,\tilde{V_3}$ are the copper and gold layers of thickness
  $\Delta$
  each.}\\[12pt]

\noindent
Fig. 2 \parbox[t]{13cm}{Constraints for the constants of hypothetical
  Yukawa-type interactions following from the new measurement of the
  Casimir force \cite{10} (curve 1), from the old Casimir force
  measurements between dielelectric bodies \cite{11} (curve 2), from
  the Cavendish-type experiments \cite{14} (curve 3), Ref. \cite{15}
  (curve 4), Ref. \cite{16} (curve 5) and from E\"{o}tv\"{o}s-type
  experiments \cite{17} (curve 6). The permitted regions of $\alpha$
  and $\lambda$ lie below the curves.}

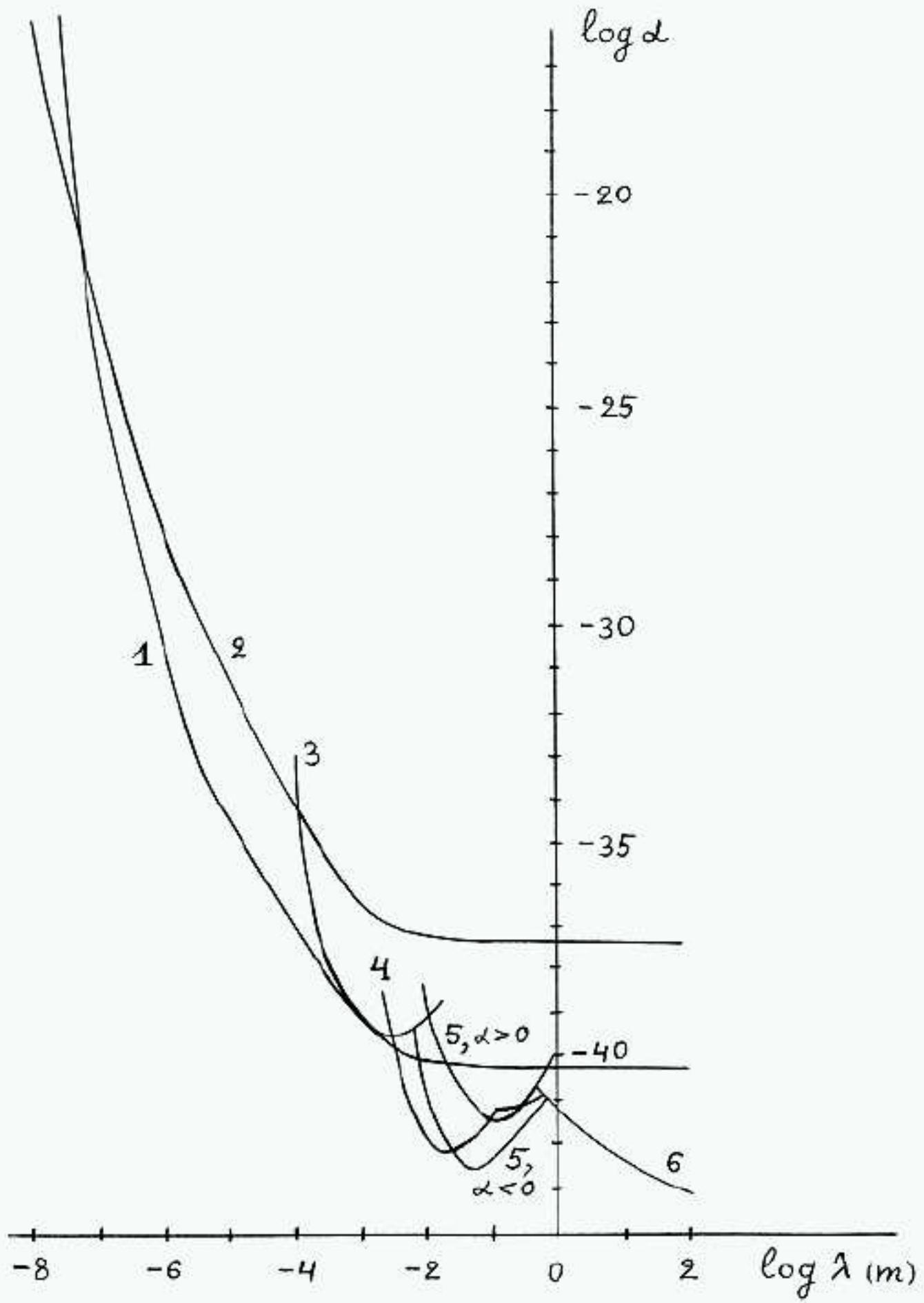
\end{document}